\documentclass{elsart}

\usepackage{natbib}

\usepackage{epsfig}

\usepackage{amssymb}

\long\def\comment#1{}
\def\Dvect#1{\mbox{\boldmath $#1$}}
\def\myfigwidth{0.97\textwidth}

\begin{document}

\begin{frontmatter}

\title{Dynamical transport of asteroid fragments from the $\nu_6$ resonance}

\author[Ito]{Takashi Ito} {\and}
\ead{\sf ito.t@nao.ac.jp}
\author[Renu]{Renu Malhotra}
\address[Ito]{Center for Computational Astrophysics,
              National Astronomical Observatory of Japan,
              Osawa 2--21--1, Mitaka, Tokyo 181--8588, Japan.}
\address[Renu]{Lunar \& Planetary Laboratory,
               University of Arizona,
               1629 E. University Boulevard, Tucson, AZ 85721--0092, USA}

\begin{abstract}
A large disruption in the main asteroid belt can cause a large flux,
an ``asteroid shower'', on the terrestrial planets.
We quantitatively examine the hypothesis that such an event was the 
cause of the lunar late heavy bombardment (LHB).
We performed numerical integrations of about 20000 test particles
starting in the vicinity of the $\nu_6$ secular resonance in the main
asteroid belt. The purpose of these integrations is to calculate, for 
each of the terrestrial planets, the collision probability of asteroids
coming from an asteroid break-up event in the inner part of the main belt. 
Compared with previous studies, we simulate nearly two orders of magnitude
larger number of particles, and we include the orbital effects of the eight 
planets, Mercury to Neptune.
We also examined in detail the orbital evolution of asteroid fragments 
once they enter the Earth's activity sphere, including the effect of the 
Earth--Moon orbit.
We obtained the collision probability, the distributions of impact velocities, 
impact positions, and impact angles of asteroid fragments on the Moon. 
The collision probability with the Moon ($\sim 0.1${\%}) suggests 
that a fairly large parent body, 1000--1500 km in diameter, is required
if the LHB event is to be ascribed to a single asteroid disruption. 
An even larger parent body is required for less favorable initial
conditions than we investigated here.
We conclude that an asteroid disruption event is not a viable
explanation for the LHB.
\end{abstract}

\begin{keyword}
Asteroid \sep celestial mechanics \sep resonance \sep collision \sep crater
\end{keyword}

\end{frontmatter}

\section{Introduction}
Among the many aspects of the impact history
of the terrestrial planets related to asteroids,
the crater record on the Moon
and the intense collisional event collectively called the late heavy
bombardment (LHB) or the terminal lunar cataclysm around 4 Ga
\citep[cf. ][]{ryder90,hartmann2000} is particularly interesting.
Evidence of LHB began to accumulate when
Ar--Ar isotopic analyses of Apollo and Luna samples suggested several
impact basins of the nearside of the Moon had been produced 
during the 3.88--4.05 Ga interval \citep{tera73,tera74}.
Recent results on the impact ages of lunar meteorites also suggest
that an intense period of bombardment on the Moon occurred about 3.9 Ga, 
i.e. 0.5--0.6 billion years after the formation of the Earth--Moon system 
\citep{cohen2000,kring2002}.

The discussion on the existence or non-existence of the lunar cataclysm
is still somewhat controversial.
Meanwhile the search for the possible dynamical cause of LHB is
as important as the geological and geochemical research of the event itself.
At present, there is no good consensus on the nature or the source(s)
of the impactors.
Recent papers have suggested that LHB may have been due to a
large flux of asteroids and comets (Kuiper Belt objects) in the inner solar
system caused by the orbital migration of the outer planets 
\citep{levison2001,gomes2005}.  Another possible cause of the LHB that 
has been suggested previously --- and one that we examine in the 
present work --- is a temporary and heavy asteroid shower created by a large
collisional event in the main asteroid belt \citep{zappala98}.
\citet{zappala98} evaluated the number of
impactors produced in different size ranges by disruption events
that might have created some of the existing asteroid families. 
Their estimates
show that an impact flux from such an asteroid-disruption event, lasting 
2--30 million years in the form of asteroid showers, could be responsible 
for the lunar cataclysm.

Previous studies, including \citet{zappala98}, \citet{gladman97} and 
\citet{morbidelli98}, have estimated the collision 
probability of asteroid fragments and the timescale of its flux on 
the terrestrial planets.  
However, all these studies have been obliged to use somewhat simplified 
dynamical models, and were limited to simulations with a
relatively small number of particles (a few hundreds to a few thousands), 
mainly because of computer resource limitations.
Also, no previous study has included the orbital dynamics of the 
Earth--Moon system when estimating the collisional probability of the 
asteroid fragments on the Earth or on the Moon.
Because the typical collision probabilities with the planets are
a few percent or less, previous simulations have typically yielded
only a few colliders, and consequently their quantitative estimates 
of collisional probabilities have significant uncertainties.
As we move closer towards the goal of accurately calculating the 
collision frequency and probability of asteroid fragments on
terrestrial planets, we would like to improve the statistics by
simulating much larger numbers of particles and also assessing
the effects of the modeling simplifications in previous studies.

The purpose of this paper is to provide a better statistical result 
on the collision probability of asteroid fragments with the terrestrial
planets.
We consider asteroid fragments coming from low inclination orbits at
the $\nu_6$ secular resonance in the main asteroid belt.  We choose this 
resonance because it is presently one of the strongest resonances for the 
transport of asteroids and meteorites to the inner solar system. 
For our study, we numerically analyze the orbital evolution of more than 
20000 test particles, with initial conditions that simulate hypothetical 
asteroid disruption events in the vicinity of the $\nu_6$. 
We also performed an additional set of orbital simulations of a large 
number of test particles as they evolve within the Earth's activity sphere,
including the dynamical effects of the Earth-Moon orbit. (The initial 
conditions for the latter simulations are derived from the orbital distribution 
function of those particles originating from the vicinity of the $\nu_6$
that manage to enter the Earth's activity sphere.)
For the range of initial source locations that we consider in this study 
(low inclination orbits near the $\nu_6$),
our results provide the best estimates currently available for asteroid 
collision probabilities on each of the terrestrial planets, as well as for 
the distributions of impact velocity, impact angle and spatial distribution 
of impacts on the Moon.
%

\begin{figure}[htbp] \centering
   \epsfxsize=\myfigwidth
   \vspace{50mm}
   \epsfbox{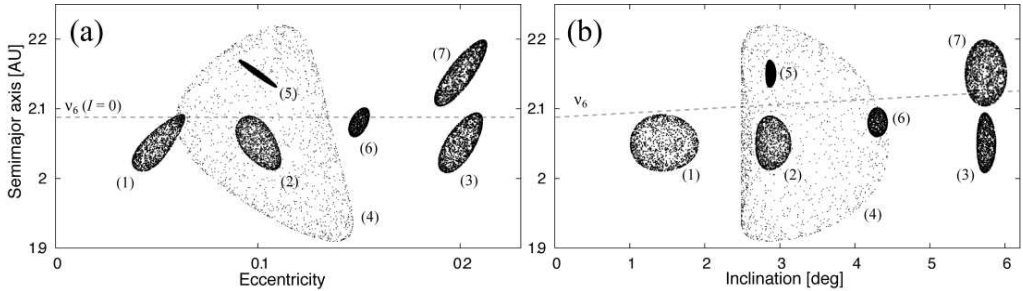}
   \caption[]{
      Initial osculating orbital elements of the asteroid
      fragments in our numerical integrations.
      (a) Eccentricity $e$ and semimajor axis $a$.
      (b) Inclination $I$  and semimajor axis $a$.
      The dashed lines in each panel indicate the
      approximate location of the $\nu_6$ resonance
      \citep[][]{morbidelli91a}.}
   \label{fig:init-eia}
\end{figure}

\section{Dynamical model and initial conditions}
We start with test particles near/in the $\nu_6$ resonance
and numerically integrate their orbital evolution under the gravitational
effect of eight major planets, from Mercury to Neptune.
The major planets are assumed to have their present masses and orbital
elements. All celestial bodies are treated as point masses dynamically,
although planetary and solar physical radii matter when we calculate
collisions between a test particle (i.e. an asteroid fragment)
and a large body.
No consideration is given to post-Newtonian gravity,
tidal force, gas drag, solar equatorial bulge, and other
non-gravitational or dissipative effects such as the Yarkovsky effect.

To emulate disruption events that create a swarm of
asteroid fragments, we assume isotropic and
equal-velocity disruptions; all fragments have the
same and isotropic initial ejection velocity, $v_0$.
We chose $v_0 = 0.1$ or 0.2 km/s, and we also tested $v_0 = 0.8$ km/s
for comparison.
We selected seven initial positions of asteroid fragments
to sample a range of locations in the vicinity of $\nu_6$
in orbital element space ($a$, $e$, $I$) as in Fig.~\ref{fig:init-eia}.
The detailed initial conditions are shown in
the upper part of Table \ref{tbl:initial}.
We created sets of orbital elements of asteroid fragments as follows:
(1) Select an appropriate point in the osculating orbital element space
    as a disruption center. This fixes six orbital elements of the
    disruption center; equivalently, six variables of three dimensional
    position and velocity of the disruption center 
    $(\Dvect{r},\Dvect{v}) = (x,y,z,v_x,v_y,v_z)$ in Cartesian coordinate.
(2) Give an ejection velocity, $\Dvect{v}_0$,
    as in Table \ref{tbl:initial}
    to each of the hypothetical asteroid fragments so that a fragment's 
    velocity becomes $\Dvect{v} + \Dvect{v}_0$ in Cartesian coordinates. 
    For each fragment, we add $\Dvect{v}_0$ with the same magnitude
    $|\Dvect{v}_0|$ with different directions using random numbers so that
    the disruption becomes ``isotropic''.
(3) Finally, convert the coordinates and velocities of the fragments,
    $(\Dvect{r}, \Dvect{v}+\Dvect{v}_0)$, into six osculating orbital
    elements; the resulting $a,e,I$ distributions are shown in 
    Fig.~\ref{fig:init-eia}.

For the currently existing asteroid families,
estimates of the initial ejection velocity of asteroid fragments
are in the range of $v_0 = 0.1$--0.2 km/s \citep{zappala96,cellino99}.
However, this is an overestimate, because account must be taken
of the Yarkovsky thermal force which works effectively on asteroid 
families to increase their velocity dispersion over time
\citep{farinella98,bottke2001}.
Thus, our adopted values of $v_0$, are actually somewhat larger than those
of the disruptions that created the observed asteroid families, but are 
appropriate for the putative breakup of a parent body as large as 
1000--1500 km in diameter, which, as we show later,
would be necessary to cause the LHB in this scenario.

For each of the seven initial disruption locations, we placed about
3000 test particles (20756 particles in total), and numerically
integrated their orbital evolution for up to 100 million years.
When a test particle goes within the physical radius of
the Sun or that of planets, we consider the particle to have
collided with that body, and remove it from the computation.
Also, when the heliocentric
distance of a test particle gets larger than 100 AU,
the particle's integration is stopped.
The ``survivors'' are particles that remain within 100 AU heliocentric distance
without colliding with the Sun or any planet for 100 Myr.
For the numerical integration we used the regularized mixed-variable 
symplectic method ({\sc swift\_{}rmvs3} by \citet{levison94}).

We should explain here the reasons why we have selected initial conditions
that are very close to the $\nu_6$ resonance. We do not necessarily think
that these are realistic initial conditions of actual asteroid disruptions
in the main belt. Large asteroids are not likely to remain close to a strong
resonance, such as the $\nu_6$, for hundreds of million years (which is the
time interval between the planet formation era and the LHB). Thus, large
collisional disruptions, if they occur, would be expected to be centered
away from strong resonances. However, the fragments that eventually collide
with the inner planets are delivered into planet crossing orbits by entering
strongly unstable resonance zones. In this manuscript our focus is
on obtaining the collision probability of asteroid fragments that are
delivered to the inner solar system via the $\nu_6$.
We use as many test particles as computationally feasible
in order to obtain the basic orbital statistics data for relevant research.
In this sense, we think our choice of initial conditions of test particles 
close/in the $\nu_6$ resonance is justified and useful for our purpose.

\begin{table}[htbp]\centering
\begin{tabular}[b]{cccccccc}
\hline
Case          & (1)  & (2)  & (3)  & (4)  & (5)  & (6)   & (7)   \\
\hline
$N_{\rm tp}$  & 2961 & 2962 & 2961 & 2967 & 2967 & 2962 & 2976  \\
\hline
$a$     (AU)  & 2.05 & 2.05 & 2.05 & 2.05 & 2.15 & 2.08 & 2.15  \\
$e$           & 0.05 & 0.10 & 0.20 & 0.10 & 0.10 & 0.15 & 0.20  \\
$I$     (deg) & 1.43 & 2.87 & 5.73 & 2.87 & 2.87 & 4.30 & 5.73  \\
$\omega$ (deg)& 330.1& 181.3& 206.5& 311.3&  81.3&  35.9& 351.2 \\
$\Omega$ (deg)& 149.8& 103.7& 192.7& 114.7& 121.0& 103.7& 235.8 \\
$l$      (deg)&  55.5& 102.6&  56.0&  97.9& 205.4&  66.5&  46.5 \\
\hline
$v_0$ (km/s)  & 0.2  & 0.2  & 0.2  & 0.8  & 0.1  & 0.1  & 0.2   \\
\hline
Sun (\%)      & 66.0 & 71.6 & 73.1 & 47.3 & 52.8 & 75.4 & 65.5  \\
Mercury (\%)  & 1.01 & 0.68 & 1.38 & 0.57 & 0.37 & 0.84 & 0.97  \\
Venus (\%)    & 6.11 & 5.06 & 4.56 & 3.24 & 2.90 & 5.00 & 3.83  \\
Earth (\%)    & 4.42 & 3.17 & 2.57 & 2.90 & 2.33 & 3.24 & 2.96  \\
Mars (\%)     & 0.71 & 0.64 & 0.54 & 0.88 & 0.94 & 0.20 & 0.94  \\
Jupiter (\%)  & 0.91 & 0.57 & 0.27 & 0.61 & 0.30 & 0.61 & 0.94  \\
Saturn (\%)   & 0.03 & 0.03 & 0    & 0    & 0.03 & 0.07 & 0     \\
$> 100$AU (\%)& 14.1 & 13.0 & 11.6 & 9.30 & 10.8 & 12.7 & 15.8  \\
survivors (\%)& 5.44 & 4.02 & 4.66 & 34.1 & 28.4 & 0.71 & 8.27  \\
\hline
\end{tabular}
\caption[]{The number of test particles $N_{\rm tp}$,
  osculating orbital elements $(a, e, I, \omega, \Omega, l)$
  of each disruption center, ejection velocity $v_0$,
  and the collision probability of asteroids that hit the Sun and planets
  in our numerical integrations.
  The fraction of particles that went beyond 100~AU and that of the particles
  that have survived over 100 million years are also shown.
  No collisions with Neptune or Uranus were observed in our simulations.
}
\label{tbl:initial}
\end{table}

\begin{figure}[htbp] \centering
  \epsfxsize=\myfigwidth
  \vspace{50mm}
  \epsfbox{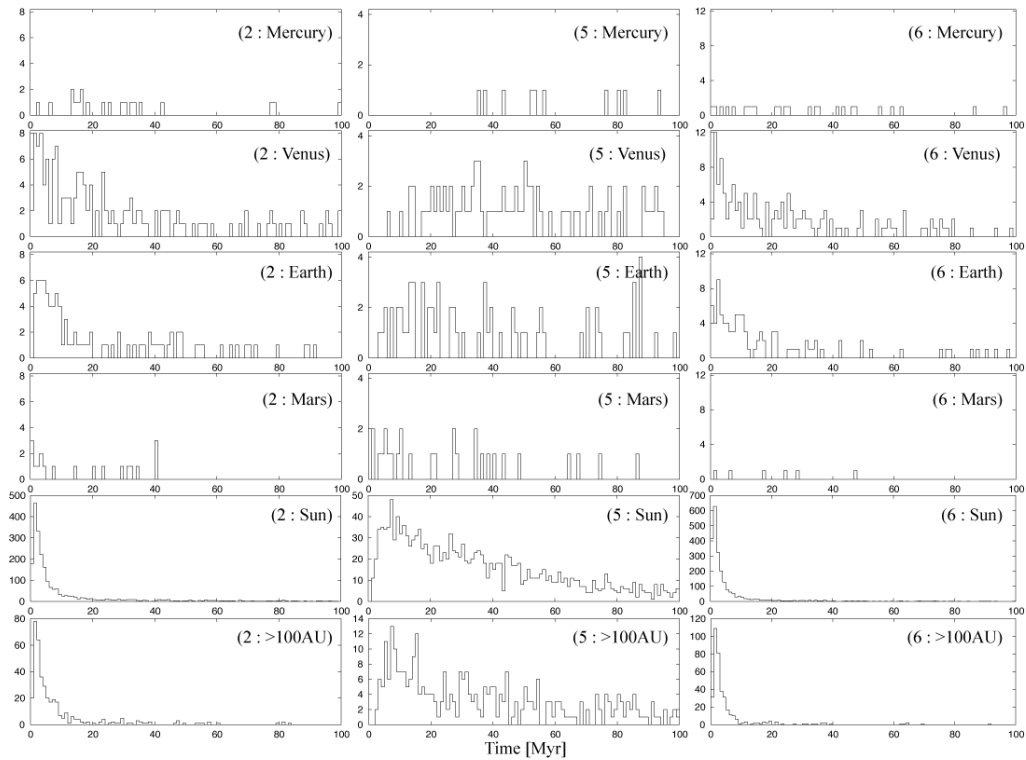}
  \caption[]{The number of particles that collided with the terrestrial
       planets and the Sun, and that went beyond 100 AU
       starting from the initial conditions (2; left panels),
      (5; middle panels) and (6; right panels).}
  \label{fig:colhist}
\end{figure}

\section{Asteroid flux on terrestrial planets}
The lower part of Table \ref{tbl:initial} summarizes the collision
probability of test particles on the planets and on the Sun, for each of
the simulations.
Except for the cases (4) and (5),
approximately 70\% of the particles collided with the Sun.
About 10--15\% of the particles were removed by going further away
than 100 AU.
%
In our numerical model, the typical dynamical behavior of particles coming
from the $\nu_6$ that eventually hit the terrestrial planets and the Sun 
as well as those particles that are ejected from the system due to scattering
by Jupiter, is qualitatively the same as has been demonstrated by
previous numerical studies as \citet{gladman97} or \citet{morbidelli98}.
Because our numerical integrations are focused on the $\nu_6$
with many more particles than the previous studies, and also because we 
include Mercury in our numerical model, we could observe collisions with
Mercury and Saturn (see Table \ref{tbl:initial}), whereas \citet{gladman97}
did not observe any collisions on these two planets. This is one illustration
of the improved collision statistics that our calculations provide.

As shown in Table \ref{tbl:initial}, several per cent of asteroid fragments
eventually hit the planets, mostly the terrestrial planets.
As described in the next section, the average collision velocity of the 
fragments with Earth is about 15--20 km/s. Rough average collision velocities
with the other terrestrial planets are
  about 14 km/s for Mars,
  about 19 km/s for Venus, and
  about 32 km/s for Mercury.
Fig.~\ref{fig:colhist} shows examples of the collision flux with time.
Although some of the panels in Fig.~\ref{fig:colhist} have small number 
statistics, some systematic trends are noticeable. 
The peak of collision flux on the planets
occurs first at Mars, and progressively later at Earth, Venus and Mercury.
(This trend is most noticeable in the middle panels, for the case (5)
in Fig.~\ref{fig:colhist}.)
The peak of the collisions with the Sun, as well as the peak of the flux
of particles that go too far away ($>100$ AU), occurs even earlier than 
the peak of the collisions with the terrestrial planets.  Generally in 
our calculation, $e$ and $I$
of the particles that are close to the $\nu_6$ center
are pumped up quickly by $\nu_6$ itself and by an associated 
Kozai oscillation,
%
%
in less than a few to ten million years \citep[cf. ][]{michel96,gladman2000}.
Many of the particles whose eccentricities reach very high values
either directly hit the Sun, or have a close encounter with Jupiter;
in the latter case, they are usually scattered outward and are
eventually eliminated from the system.
This explains both the rapid production and the rapid decay rate of the 
solar colliders and ``too-far'' particles.
On the other hand, the particles excited to only moderate eccentricity  
evolve by means of many close encounters with the terrestrial
planets which gradually reduce their semimajor axes, making them
migrate within the terrestrial planetary zone on a longer timescale.
These particles are the most common candidates for planetary collisions.

The timescale of asteroid flux on the planets strongly depends on
the initial location of the fragments and on their initial ejection 
velocity \citep{gladman97,morbidelli98,morbidelli99}.
When the initial distribution of asteroid fragments is not so widely scattered
(i.e. when their $v_0$ is small) but the location
of the disruption event is far away from the resonance center, such as case (5),
it takes several million years for particles to achieve planet-crossing orbits,
and the flux of solar colliders, too-far particles, and planetary colliders 
lasts for many tens of millions of years
(the middle panels of Fig. \ref{fig:colhist}). This is somewhat in contrast 
with the cases of those initial conditions whose center is closer to the 
$\nu_6$ (e.g., cases (2) and (6), Fig. \ref{fig:colhist}): In these cases
the flux of solar colliders and too-far particles decays rapidly, with
a timescale of 10 million years, although the flux of planetary colliders 
lasts for many tens of millions of years.
We see many survivors in the cases (4) and (5) at 100 myr 
(Table \ref{tbl:initial}).
In contrast, when $v_0$ of asteroid fragments
is small and the location of disruption event is near the resonance
center, such as the case (6), the removal efficiency of asteroid fragments
is very high, and the impact flux decay timescale is very short
as we see in the right panels of Fig. \ref{fig:colhist}.

Table \ref{tbl:initial} shows that the collision probability of asteroid
fragments on Earth is about 3\%.
Note that we have assumed relatively low initial inclinations for the
asteroid fragments; the collisional probability would be lower for higher
initial inclinations \citep[cf. ][]{morbidelli98}.
The integrated collision probability for Venus over 100 myr is 
1.2--1.7 times larger than that for the Earth.
Also, the collision probability for Mercury is somewhat larger
than that for Mars (except perhaps in cases (4) and (5) where more 
planetary collisions would have taken place if we continued the integrations).
We can interpret this trend as the combined effect of planetary physical
cross sections and average relative velocity of asteroid fragments
at each of the planetary orbits, assuming the particle-in-a-box approximation
stands. But we need simulations with larger numbers of particles to get more 
reliable statistics for the differences in collision rates amongst the
terrestrial planets.
Previous research has used even fewer number of
particles in this kind of numerical simulation: For example,
\citet{gladman97} used only 150 particles for the experiment of the
$\nu_6$ resonance, and the number of particles that
\citet{morbidelli98} used was about 400 at $\nu_6$.

\section{Collisions with Earth and Moon}
To obtain the collision probability of asteroid fragments with the Moon
and to compare it with the lunar crater record, we need to calculate the actual
impact flux on the lunar surface. However, looking at our
current numerical result described in Fig. \ref{fig:colhist}, the number
of planetary collisions is relatively small. 
For example, starting with a swarm of 3,000 particles,
we have about a hundred collisions on Earth over 100 myr.
Moreover, the collision probability of asteroid fragments
on the Moon in our numerical model
would be much smaller than that on the Earth due to the smaller
lunar gravity and physical radius, yielding only a few collisions over 100 myr.
This poor yield of planetary colliders is the source of the relatively poor
statistics from such dynamical models.

\begin{figure}[htbp] \centering
  \epsfxsize=\myfigwidth
  \vspace{50mm}
  \epsfbox{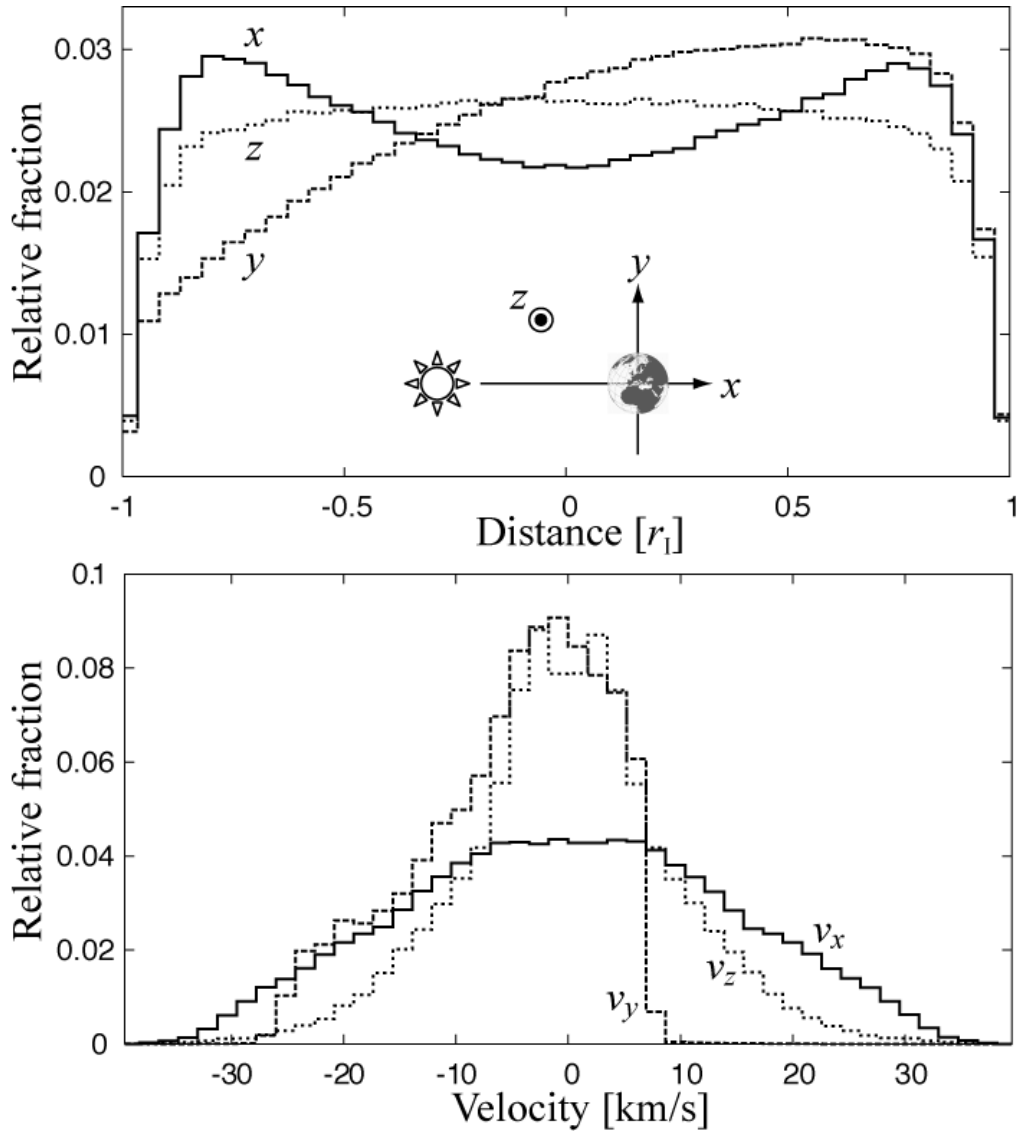}
  \caption[]{An example of the statistics at the Earth's $r_{\rm I}$
   from the case (2).
   Upper: encounter position of asteroid fragments.
   Lower: encounter velocity.
   The coordinates $x,y,z$ are centered on the Earth;
   $x$-axis is oriented from the Sun toward the Earth,
   $y$-axis is oriented toward the direction that the Earth goes, and
   $z$-axis is normal to the Earth's orbit and the north is
   defined as positive
   (see the schematic illustration in the upper panel).
   All statistics are integrated over 100 myr.
}
  \label{fig:rvI}
\end{figure}

However there is a method to improve the collisional statistics for
the Moon: although we get only a few hundred planetary colliders in our 
simulations, we have many more encounters at the planetary activity sphere.
The activity sphere, also known as ``sphere of influence''
\citep[cf. ][]{danby92}, of the
Earth $r_{\rm I}$, which is proportional to $(m_\oplus/M_\odot)^{2/5}$ where 
$m_\oplus$ is the Earth mass and $M_\odot$ is the solar mass, is
about 144 Earth radii. In our numerical result, about a million encounters
took place at this distance in each of the simulated cases.
This number is large enough to establish a time-dependent orbital
distribution function such as $F(a,e,I,\omega,\Omega,l; t)$ to create
``clones'' of asteroid fragments in order to increase the reliability of
the collision statistics on the Earth--Moon system.
What we actually did to generate clones is to slightly change the
encounter position 
$\Dvect{r}$ and velocity $\Dvect{v}$ of each of the original asteroid fragments
so that their orbital motion in the activity sphere is a bit different, as
$\Dvect{r}_{\rm clone} = (1+\delta_r) \Dvect{r}_{\rm original}$ and
$\Dvect{v}_{\rm clone} = (1+\delta_v) \Dvect{v}_{\rm original}$ where
$|\delta_r|, |\delta_v| < 0.1$ in our model.
Repetition of this procedure produces
a large number of clones that obey nearly the same orbital distribution
function $F$ as the original particles but with slightly different paths
toward the Earth and the Moon. In Fig.~\ref{fig:rvI}, 
we show an example of the integrated
distribution of encounter position and velocity of original particles at
Earth's activity sphere that we used to create clones.

We repeated this cloning procedure 1,000 times for each set of
integrations that originally involves about 3,000 particles and about a
million encounters at $r_{\rm I}$. For each of the initial condition sets,
the cloning procedure generates
three million clones with a billion encounters
around Earth's activity sphere. Using these cloned particles,
we performed another set of numerical integrations with the Sun, the Earth,
the Moon, and cloned test particles within the Earth's activity sphere.
Here we do not include the effect of other planets, but we
do include Moon's gravity. All the clones start from just outside of Earth's
activity sphere, and all of them go through the sphere until they hit
the Earth or the Moon or go out of the sphere. We used the
present orbital elements of the Moon, with the true anomaly
randomly chosen from 0 to $2\pi$.
As for the numerical method, we use the regularized mixed-variable 
symplectic method again with a stepsize of 168.75 seconds (= $2^{-9}$ days).
We use geocentric frame for this calculation.

\begin{table}[htbp]\centering
\begin{tabular}[b]{crrrrrc}
\hline
 & \multicolumn{1}{c}{$N_{\rm tp}$}
 & \multicolumn{1}{c}{$N_{\rm enc}$}
 & \multicolumn{1}{c}{$n_{\rm c,E}$}
 & \multicolumn{1}{c}{$n_{\rm c,M}$}
 & \multicolumn{1}{c}{$P_{\rm c,M}$}
 & \multicolumn{1}{c}{$\frac{n_{\rm c,E}}{n_{\rm c,M}}$} \\
\hline
(1)& 2961 & 1142636 &  87486 & 3708  & 0.125\% & 23.6 \\
(2)& 2962 & 1176793 & 101766 & 4160  & 0.140\% & 24.5 \\
(3)& 2961 &  982652 &  81359 & 3618  & 0.122\% & 22.5 \\
(4)& 2967 &  777056 &  72613 & 2801  & 0.094\% & 25.6 \\
(5)& 2967 &  648519 &  53647 & 2388  & 0.080\% & 22.5 \\
(6)& 2962 &  998867 &  82014 & 3501  & 0.182\% & 23.4 \\
(7)& 2976 &  758500 &  66163 & 2840  & 0.095\% & 23.3 \\
\hline
\end{tabular}
\caption[]{Collisions of clones on the Moon.
$N_{\rm tp}$ is the number of original particles that is also shown in
  Table \protect{\ref{tbl:initial}}.
$N_{\rm enc}$ is the number of encounters of the original particles with
  Earth's activity sphere, $r_{\rm I}$.
These numbers are multiplied by 1,000 to generate clones.
$n_{\rm c,E}$ and $n_{\rm c,M}$ are the numbers of collisions of clones
  with the Earth and with the Moon.
$P_{\rm c,M}$ is the collision probability of clones to the Moon
 ($P_{\rm c,M}/100 \equiv n_{\rm c,M}/1000N_{\rm tp}$).
}
\label{tbl:stat-clones}
\end{table}

The resulting collision statistics is summarized in Table
\ref{tbl:stat-clones}. Now we have about $10^5$ collisions on the Earth
$(n_{\rm c,E})$ and a few thousand on the Moon $(n_{\rm c,M})$.
Overall collision probability of the asteroid fragments on the Moon
$(P_{\rm c,M})$ is about 0.1{\%}, and the number ratio between the
collisions on the Earth and on the Moon $( n_{\rm c,E}/n_{\rm c,M} )$
is about 22--24, which is similar to an analytical estimate \citep{zahnle97}.

\begin{figure}[htbp] \centering
  \epsfxsize=\myfigwidth
  \vspace{50mm}
  \epsfbox{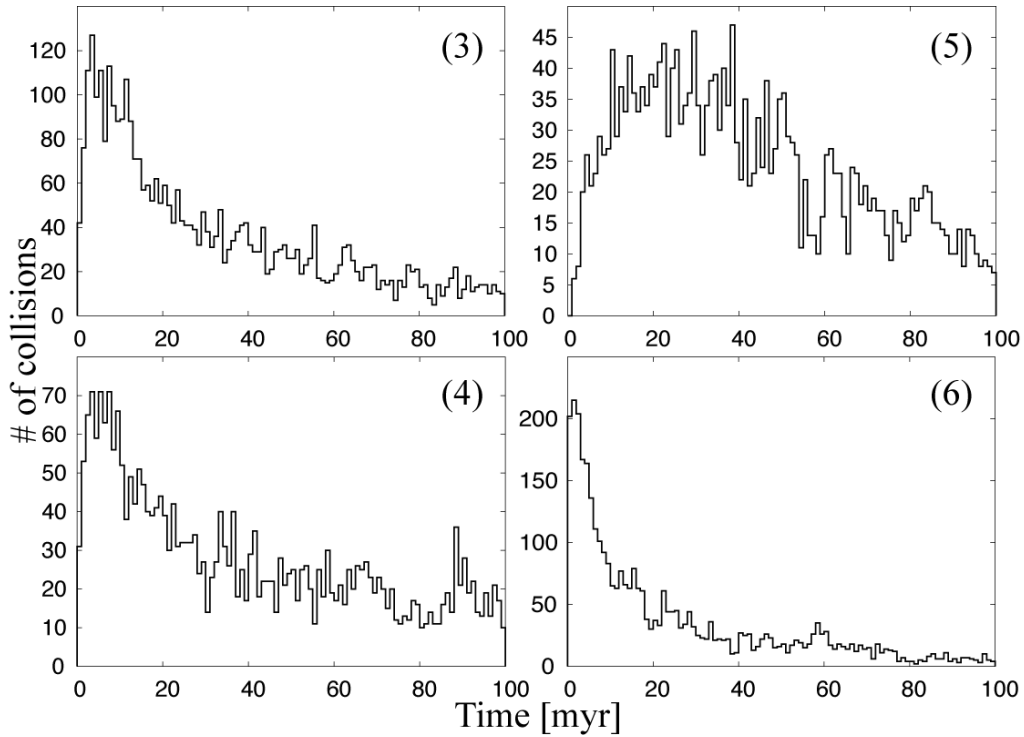}
  \caption[]{Examples of the asteroid impact flux on the Moon
    when we used the initial condition sets (3)(4)(5) and (6).}
  \label{fig:col_M}
\end{figure}

\begin{figure}[htbp] \centering
  \epsfxsize=\myfigwidth
  \vspace{50mm}
  \epsfbox{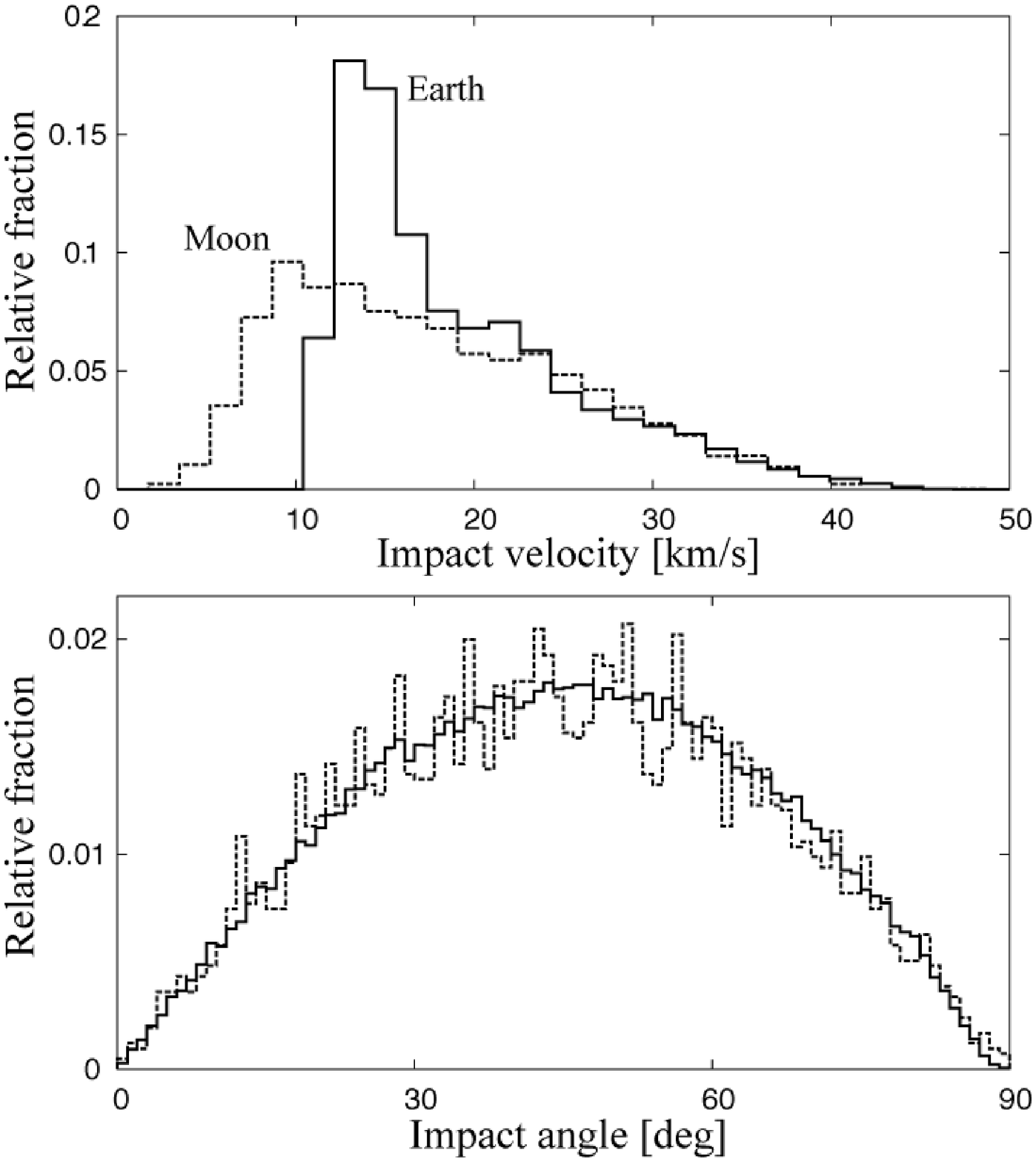}
  \caption[]{An example of the distribution of the impact velocity (the
   upper panel) and angle (the lower panel) of clones in the case (2).
   In the lower panel, the solid line is for the Earth, and the dashed
   line is for the Moon.
}
  \label{fig:vrth}
\end{figure}

Not surprisingly, the time-evolution of the 
impact flux of clones strongly depends on the initial location of 
asteroid fragments (Fig.~\ref{fig:col_M}). Different initial conditions
produce a variety of impact decay rates. However, the time-integrated
statistics of the impact velocity, the impact angles, and spatial
distribution of impacts on the Earth or the Moon are very similar for 
all initial condition sets; we show the results for one initial condition 
set (case 2) in Fig~\ref{fig:vrth}.
This result indicates that the time-integrated characteristics
of the impacts on Earth or Moon are not strongly sensitive to the initial 
velocity dispersion of the asteroid fragments or their distance from the 
resonance.
However, we cannot rule out a dependence of these distributions on
initial inclination, since all our initial condition sets had similar
low inclinations.


The variety of impact decay rates and the
distribution of impact velocity and angles obtained here can be used in 
future work to create a synthetic
crater record of the Moon to compare with the actual
lunar craters. We have also calculated the spatial
distribution of the impacts on the lunar surface. 
(Note the leading/trailing asymmetry evident in Fig~\ref{fig:rvI}.)
This can be used for
comparison with the geographical data of the lunar crater distribution
which exhibits global asymmetries \citep{morota2003,morota2005}.
Detailed comparison will appear in future publications.

\section{Summary and Conclusions}
We have explored the dynamical evolution of test particles with initial
conditions near the $\nu_6$ resonance in order to simulate the orbital
evolution of fragments from hypothetical asteroid break-up events.
Compared with previous studies such as \citet{gladman97} or
\citet{morbidelli98}, our simulations follow
nearly two orders of magnitude larger number of particles.
We calculated the collision 
probabilities on each of the terrestrial planets and on the Sun, and the 
dynamical lifetimes of asteroid fragments.
Our improved statistics shows that the collision probability of asteroid
fragments on the Earth is about 3{\%} and on Venus is about 4--6{\%},
not strongly sensitive to initial conditions in the vicinity of 
the $\nu_6$ resonance in the low inclination part of the inner asteroid belt
(Table \ref{tbl:initial}).
These numbers are slightly different from the numerical results presented
in \citet{gladman97} that had yielded the collision probability of
test particles from the $\nu_6$ region on the Earth as 5.5{\%} and
that on Venus as 1.8{\%}.

We also performed a set of orbital simulations of billions of cloned
asteroid fragments inside of Earth's activity sphere, including the
dynamical motion of the Moon using the orbital distribution function of
the particles calculated by the long-term orbital integration described above.
From these simulations, we have obtained the collision probability, impact 
velocity, impact positions and impact angles of asteroid fragments on the 
Moon.  These will be useful in a future study to compare with the observed
lunar crater record.  So far our simulations have only tested the 
particle dynamics with the current lunar orbit.
But at 4 Ga the Moon could have been closer to the Earth \citep[cf. ][]{touma94}.  
The higher lunar orbital velocity and the greater gravitational focusing effect
due to the lunar proximity to the Earth could both produce a dynamical effect
on the collision probability of asteroids.
Models of lunar evolution suggest that the Moon would experience
a large increase in its orbital semimajor axis very early in its history
and then gradually move away from the Earth over its remaining history;
the distance and orbital velocity of the Moon with respect to Earth at the 
time of the LHB depend upon model parameters (tidal dissipation functions).
This problem will be explored in a future study, as it might help to test
models of lunar evolution. 

Regarding the collision probability of the asteroid fragments on the Moon,
we obtained a small value, $\sim 0.1${\%}. The total mass of the
LHB impactors on the Moon is estimated as $\sim 0.01${\%} lunar mass
\citep{hartmann2000}. Hence, if we ascribe LHB to a single
asteroid disruption in the vicinity of the $\nu_6$ resonance, 
we would need a very large parent body, as large as 1000--1500 km
in diameter. (Moreover, the parent body would be required to be
very close to the $\nu_6$ resonance and to disrupt while it was in this 
short-lived orbit).
This size should be considered a lower limit, because the collision
probabilities would generally be lower for asteroid fragments from 
initial orbits of higher inclination or further from the $\nu_6$ or from
weaker resonances than the initial condition sets that we have simulated.
Considering the fact that in the present main belt there is no 
asteroid larger than 1000 km diameter, the required size might seem 
rather improbable.  We might need to consider the possibility that two 
or more asteroid disruptions happened at the time of LHB.
Another argument against the likelihood of the large asteroid breakup
hypothesis was recently presented \citet{strom2005}
that showed that the LHB crater size distribution is quite consistent with
the size distribution of asteroidal projectiles from the main asteroid belt.  
The main belt asteroids have a complex and distinct size distribution
\citep[e.g. ][]{jedicke98,ivezic2001,yoshida2003}
probably as a result of a collisional cascade
\citep{cheng2004,bottke2005a},
and it is quite different from what would be expected for the fragments
in a single collisional disruption \citep[e.g. ][]{michel2004}.
Both arguments above suggest that we should reject the
large-asteroid-disruption hypothesis as a mechanism for the LHB.

The more likely cause of LHB may lie in the sweeping of strong 
resonances through the asteroid belt due to the orbital migration of giant planets
\citep{levison2001,gomes2005}.
In such a mechanism, the $\nu_6$ secular resonance would be amongst the most 
powerful for causing asteroids to be excited into terrestrial 
planet-crossing orbits.  The results presented here on the dynamics of 
asteroid impactors emerging from the vicinity of the $\nu_6$ could be
useful for the further study of this possible mechanism of the LHB.
We note that in the present paper we have investigated only the dynamical 
effects of a stationary resonance, not a migrating resonance. 
Collision probabilities and impact velocities of asteroids might be
different, depending on whether the resonance region is fixed (stationary)
or migrating (sweeping). This is because in the vicinity of a resonance,
there is a large range of timescales for the excitation of eccentricity,
and there are differing amounts of phase space volume with the short versus 
long timescales.  The short timescale zone enhances particle eccentricities
quickly, so the impact velocity of an asteroid (with the planets and the Moon) 
will be higher, but the impact probability will be smaller. On the other hand, 
long timescale zone around a resonance is a source of asteroids that hit the
planets with smaller impact velocity but with a higher probability because 
the gravitational focusing is more effective at lower velocity encounters.
When resonances migrate due to planetary migration, the relative importance
of the short timescale and the long timescale resonance zones would depend 
upon the speed of planetary migration.
Future modeling of this mechanism is needed, especially the comparison
of collision probabilities and impact velocities of asteroids of
these two cases.  It would also be important to obtain improved direct estimates 
the time duration of the LHB because it is directly related to the speed of
resonance migration. For this purpose, more accurate chronology of lunar impact 
craters would help to constrain the dynamical mechanisms.
This requires further lunar exploration, including sample-return missions.

\ack
The authors are grateful to the two referees of this paper,
Alessandro Morbidelli and Bill Bottke, who suggestions 
substantially improved the quality of the manuscript.
This study is supported by the Grant-in-Aid of the Ministry of Education
of Japan (16740259/2004--2005) and by NASA research grants NNG05GI97G 
(NASA-Origins of Solar Systems Research Program) and 
NNG05GH44G (NASA-Outer Planets Research Program).





\end{document}